\documentclass[preprint2]{aastex}
\usepackage{graphicx}

\begin{document}

\slugcomment{Submitted to the Astrophysical Journal}

\shorttitle{Roberts et al.}
\shortauthors{Pulsar Wind Nebula in G11.2$-$0.3}

\title{The Pulsar Wind Nebula in G11.2$-$0.3}

\author{
Mallory S. E. Roberts,\altaffilmark{1,2}
Cindy R. Tam,
Victoria M. Kaspi,\altaffilmark{1,3}
Maxim Lyutikov,\altaffilmark{1,4}}
\affil{Dept. of Physics, Rutherford Physics Building,
McGill Univ., 3600 University Street, Montreal, Quebec,
H3A 2T8, Canada}
\author{Gautam Vasisht}
\affil{Jet Propulsion Lab., Calif. Institute of Technology, 4800 Oak Grove Drive, Pasadena, CA 91109}
\author{Michael Pivovaroff}
\affil{Space Sciences Lab., Univ. of California, Berkeley, 
260 SSL \#7450, Berkeley, CA 94720-7450}
\author{Eric V. Gotthelf}
\affil{Columbia Astrophysics Lab., Columbia Univ., 550 West 120th Street, New York, NY 10027}
\author{Nobuyuki Kawai\altaffilmark{5}}
\affil{Dept. of Physics, Tokyo Institute of Technology,
2-12-1 Ookayama, Meguro-ku, Tokyo 152-8551, Japan}

\altaffiltext{1}{Dept. of Physics and Center for Space Research,
M.I.T., Cambridge, MA 02139}
\altaffiltext{2}{Quebec Merit Fellow}
\altaffiltext{3}{Alfred P. Sloan Research Fellow, Canada Research Chair}
\altaffiltext{4}{CITA National Fellow}
\altaffiltext{5}{RIKEN (Institute of Physical and Chemical Research),
2-1 Hirosawa, Wako, Saitama, 351-0198 Japan}

\begin{abstract}
We present an X-ray and radio study of the wind nebula surrounding the
central pulsar PSR J1811$-$1925 in the supernova remnant G11.2$-$0.3.
Using high resolution data obtained with the {\it Chandra} X-ray
observatory and with the VLA radio telescope we show
the X-ray and radio emission is 
asymmetric around the pulsar, despite the latter's central position in the
very circular shell. 
The new X-ray data
allows us to separate the synchrotron emission of the pulsar wind
nebula from the surrounding thermal emission and that from the pulsar itself.
Based on X-ray
data from two epochs, we observe temporal variation of the location of
X-ray hot spots near the pulsar, indicating relativistic motion. 
We compare thermal emission observed  
within the shell, which may be associated with the forward shock
of the pulsar wind nebula, to thermal emission from a nearby portion of the 
remnant shell, the temperature of which implies an expansion velocity
consistent with the identification of the remnant with
the historical event of 386 A.D. 
The measured X-ray and radio spectral indices of the nebula synchrotron 
emission
are found to be consistent with a single synchrotron cooling break. The 
magnetic field implied by the break frequency 
is anomalously large given the apparent size and age of the nebula if
a spherical morphology is assumed, but is consistent with a 
bipolar morphology. 

\end{abstract}


\section{Introduction}
The overall structure and evolution of the X-ray emission produced by a 
spherically symmetric relativistic particle
wind coming from a young pulsar in a supernova remnant has been 
modelled extensively \citep{rc84,kc84,vand01,bcf01}. 
In addition, some progress has been made
on the fine structure and time dependence of the sites of shock acceleration
in the inner regions \citep{ga94}. 
The observational tests of these models have
been dominated by discussions of the Crab nebula. The advantages of
studying the Crab, with its high luminosity, known age, and well observed
pulsar, are many. However, the Crab is less than an ideal test
case for understanding pulsar wind nebulae (PWNe) in supernova
remnants (SNRs) as a class due to its anomalously high luminosity
and lack of an observable SNR shell.
There are a number of observed radio composite SNRs which consist of a shell
and a spectrally distinct inner nebula, presumably PWNe. Only in
a few cases, however,  has the central pulsar been detected (eg. W44,
see Kaspi \& Helfand 2002\nocite{kh02} for a review, also 
Gotthelf 2001\nocite{got01}), 
leaving the
spin-down energy driving the emission to be inferred from the PWN itself.   

G11.2$-$0.3 is a remarkably spherical young SNR 
plausibly associated with the historical event of 386 A.D. 
recorded by Chinese astrologers \citep{cs77}. A 65~ms pulsar with 
spin-down energy
$\dot E=6.4\times 10^{36}\,{\rm ergs}\,{\rm s}^{-1}$ 
was discovered in X-rays by {\it ASCA} \citep{tori97,tori99}.
The characteristic spin-down age is much greater than the apparent age of 
the SNR and that inferred from its remarkably central position within the 
remnant given any reasonable transverse velocity (Kaspi et al. 
2001\nocite{kasp01}, hereafter
Paper I). The simplest explanation for this apparent age 
discrepancy is that the initial spin period of the pulsar is very near the 
current spin period, which in turn suggests that $\dot E$ has remained 
nearly constant since the supernova explosion. The distance to G11.2$-$0.3 has 
been reasonably
well determined to be $d\sim 5~$kpc from HI measurements \citep{bmd85,gree88}. 
We therefore have a system
where the assumptions of spherical symmetry, at least in regards to
the shell around the pulsar,  and constant known energy input
are observationally supported, and reasonable estimates of the physical
size of various components can be made. 

The bright spherical radio shell, reminiscent of the remnants of
Tycho's and Kepler's supernovae \citep{down84}, and the 
lack of an obvious hard central component in the $ROSAT$ PSPC image,
initially led some authors to conclude that G11.2$-$0.3 was the remnant
of a Type Ia supernova \citep{rey94}. 
The first hints of a central plerionic component came from high
frequency single dish radio measurements which suggested a central flattening
of the radio spectrum \citep{mr87}. 
{\it ASCA} observations first clearly demonstrated 
that there was a central, non-thermal X-ray source \citep{vasi96}. 
This component was
resolved by $Chandra$, the image of which has been presented in Paper I. 
Further single dish radio observations have clearly demonstrated
the existence of a flat central radio component \citep{kr01}. 
Archival VLA observations
separate the central plerion from the shell and measure
the spectral index to be $\alpha_r\sim 0.25$ (defined as 
$S=S_r(\nu/\nu_r)^{-\alpha_r}$, where
$S_r$ is the energy flux density at frequency $\nu_r$), a fairly typical value
(Tam, Roberts, \& Kaspi 2002,\nocite{trk02} hereafter Paper II). 
In this paper we combine the radio data with the $Chandra$ data to make 
spectral measurements 
of the pulsar itself, the hard pulsar wind nebula, and soft thermal 
emission possibly related to the PWN which we compare to a portion of the shell.
We also present evidence
for apparent temporal variation in the positions of emission features in 
the central portion of the X-ray PWN, 
and make a detailed comparison of the X-ray morphology with a new 3.5~cm VLA
image. We then compare
these observations to models of PWN evolution and explore the 
possibility that within the SNR shell a wind termination shock produces the 
variable hard emission, and the forward shock of the PWN expanding into  
the stellar ejecta produces the interior thermal structures. 

\section{Observations and Analysis}

\subsection{X-Ray and Radio Observations}

NASA's {\it Chandra X-ray Observatory} observed G11.2$-$0.3 at two
epochs, the first (Sequence Number 50076) on 2000 August 6, and
the second (Sequence Number 50077) on 2000 October 15, for 20~ks
and 15~ks respectively. 
The remnant was positioned on the back-illuminated CCD chip S3 of
the ACIS instrument in standard exposure mode with a
time resolution  of 3.2~s, too coarse to detect pulsations from
the pulsar.
The data were reduced and analysed with CIAO 2.2.1 using Caldb 2.12, XSPEC 
V.11.1.0 and the MIRIAD 
software packages. 
Details of the image analysis and a three color X-ray image of the SNR can be
found in Paper I. Here, we focus on the emission interior to the
shell. 

Analysis of archival 20~cm and 6~cm VLA data presented in Paper II 
revealed the extent of the radio PWN by showing 
that most of the central enhancement has a significantly
flatter spectrum than the surrounding shell. In order to further enhance the
relative brightness of the PWN over the shell, we have obtained 3.5~cm data 
with the VLA in the DnC, C, and CnB configurations as part of a campaign to
better constrain the broad-band radio spectrum and measure the shell
expansion. Details of the radio observations will be presented elsewhere
(Tam et al. in preparation). 

\subsection{Spatial Analysis}

\begin{figure}
\begin{center}
\includegraphics[width=15cm,keepaspectratio='true']{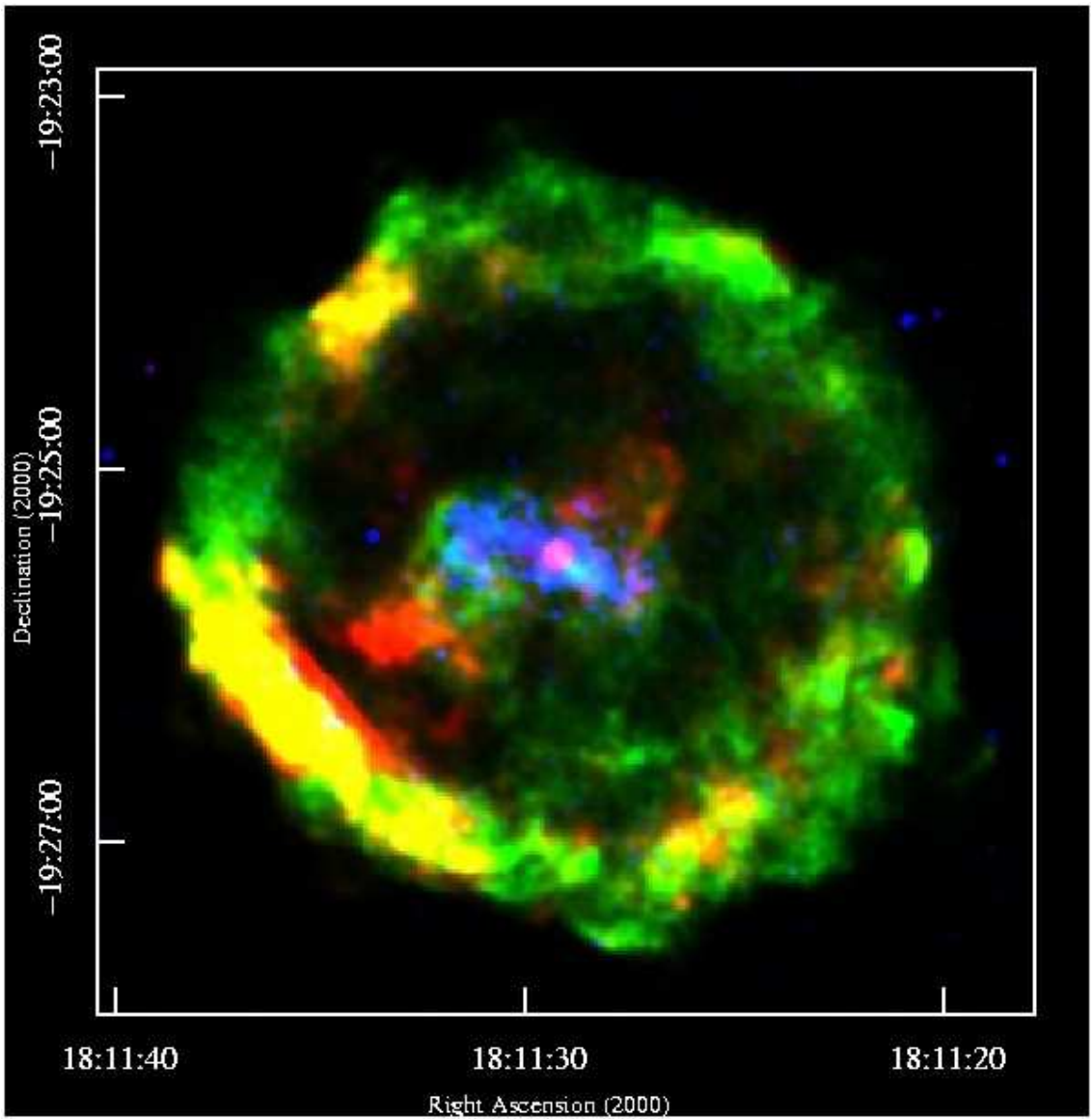}
\end{center}
\figcaption{\label{fig:multi}
Comparison of X-ray PWN emission and radio emission.
Red represents photons of energies 0.6--1.65~keV, green is 3.5~cm radio,
and blue is 4-9 keV photons.
}
\end{figure}

The multi-wavelength image of G11.2$-$0.3 is presented in 
Figure~\ref{fig:multi}. The 3.5~cm radio image is shown in green,
the soft (0.6-1.65~keV) X-ray image, dominated by thermal emission, is 
in red, and the hard (4-9~keV) X-ray image is in blue.
This image shows the relationship between the interior thermal X-ray emission, 
the radio PWN and the bright, hard X-ray emission of the PWN. It also shows
how the relative brightness of the soft X-ray emission and the radio 
emission changes around the shell. The point source at the center is the pulsar,
which is only seen in the X-ray images.
Paper I noted that the hard X-ray PWN
consists of a bright region southwest of the pulsar, a narrow, almost
linear component to the northeast, and a larger, faint, diffuse component 
with reasonable bi-lateral 
symmetry around the pulsar (see Figure~2 in Paper I). 

\begin{figure}
\plotone{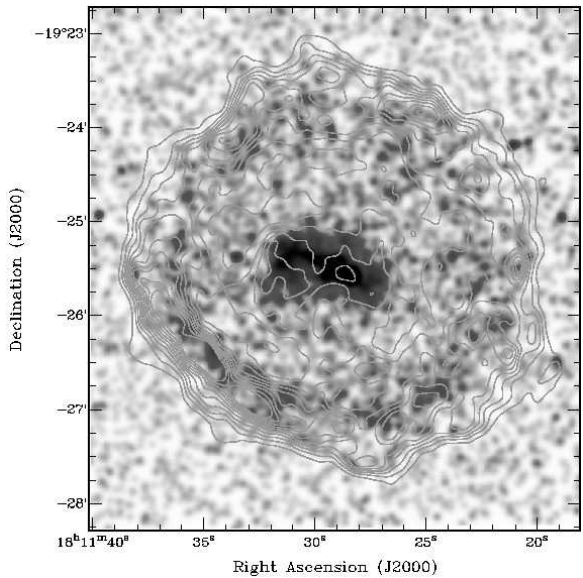}
\figcaption{\label{fig:hardrad}
Comparison of hard X-ray emission and radio emission.
Greyscale is 4-9 keV X-rays, contours are 6~cm radio emission.
}
\end{figure}

In  Figure~\ref{fig:hardrad} we show the exposure corrected, 4-9 keV summed 
image of the two observing epochs, smoothed with a $5^{\prime \prime}$ 
Gaussian, with 6~cm radio (which better shows a combination of 
flatter and steeper features) contours overlain.
The SNR shell predominantly emits soft, thermal X-ray emission which
is mostly invisible above 4 keV, and so the features in this image show
regions of non-thermal emission. The central PWN 
is quite hard, and the greyscale levels in this image 
have been chosen to highlight the
difference between the bright and faint parts of the PWN.
The extent and shape of the radio PWN correlate well with the faint hard PWN 
X-ray component. 
To the northeast of the pulsar, the narrow bright hard X-ray feature goes 
through a region of 
low radio luminosity, and then terminates at a bright arc in the radio emission. 
To the southwest, the bright hard X-rays coincide with the radio emission, with a radio
enhancement coincident with the X-ray peak (the `spot' region referred to below).
The shell is clearly seen in hard X-rays, although they are concentrated
towards the inner edge of the shell, especially in the bright southwestern
section. These hard X-rays are largely interior to the peak emission of the 
radio shell in this region.

\begin{figure}
\plotone{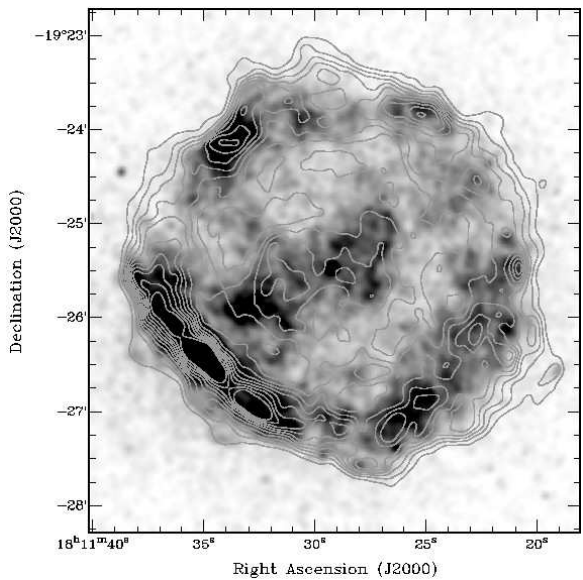}
\figcaption{\label{fig:softrad}
Comparison of soft X-ray emission and radio emission.
Greyscale is 0.6--1.65~keV X-rays, contours are 6~cm radio emission.
}
\end{figure}

In Figure~\ref{fig:softrad} we show the soft (0.6$-$1.65 keV) 
X-ray emission with the
same radio contours as in Figure~\ref{fig:hardrad}.
There is a large, soft X-ray structure which we refer to as the soft ``PWN", 
which is centered on the pulsar, reaching to the northwest and 
southeast. 
The ratio of medium to soft X-ray flux, suggests this structure does not
have any large spectral variations (see Paper I). 
This soft X-ray emission seems to
outline much of the radio PWN and largely coincides with 
radio features that were shown in Paper II 
to have steeper spectra consistent with the
radio shell emission rather than the PWN emission. 
The soft X-ray and radio shells are
quite similar, except the radio emission is slightly more extended than are
the X-rays around
most of the shell. However, the bright enhancement in the southeastern portion
of the shell seems to be broader in thermal X-rays relative to the radio than 
elsewhere in the shell.



\begin{figure}
\plotone{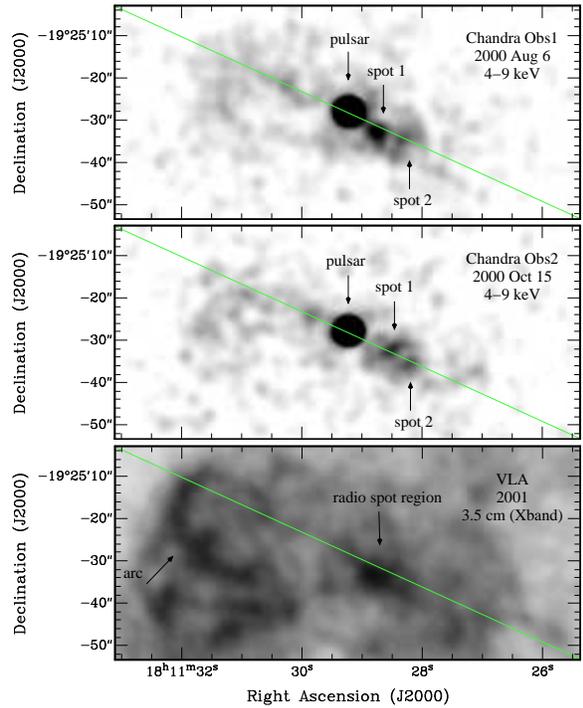}
\figcaption{\label{fig:wisps}
ACIS observing
epochs 1 (top) and 2 (bottom), smoothed with a $3^{\prime \prime}$ Gaussian, and
3.5~cm (bottom) VLA X-band image with a $3.1^{\prime \prime}\times
2.8^{\prime \prime}$ restoring beam. The line in
the images shows the 1-D slice for the profiles shown in Figure~\ref{fig:prof}.
}

\end{figure}
\begin{figure}
\plotone{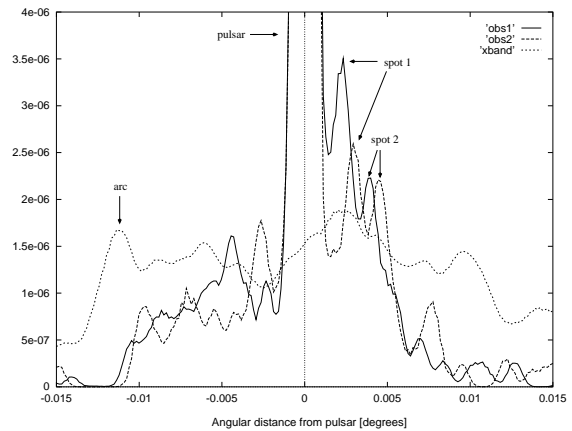}
\figcaption{\label{fig:prof} Intensity profile of a line through the spots for the 
two observing epochs and the radio image, showing the displacement
between the two epochs. The y-axis is in arbitrary units of flux density.}

\end{figure}

To study the details of the PWN,  we produced 
separate hard X-ray images for the two observation epochs 
whose coordinates we aligned by using the bright
emission of the pulsar to correct for slight differences in the
aspect solution. These were then exposure corrected and 
smoothed with a $3^{\prime \prime}$
Gaussian (Figure~\ref{fig:wisps}). For comparison, we also present
the same region from the 3.5~cm radio observation, 
which best shows the PWN due to
its flatter spectra relative to the shell.
The brightest features in both X-ray images other than the 
pulsar are two spots to the southwest of the pulsar, which are
$\sim 10\sigma$ and $\sim 7\sigma$ above the surrounding PWN level. 
To better see the relative amplitudes,  we also show line intensity
profiles of all three images of the slice indicated in the images 
(Figure~\ref{fig:prof}).
The positions of the two spots were determined for each epoch both
by centroiding and by fitting a 2-D Gaussian on top of a constant background,
the two methods giving consistent results. 
The latter method gives uncertainty estimates with which we add
in quadrature a relative aspecting uncertainty of $0.31^{\prime \prime}$
using the pulsar as our reference (see aspecting calibration memos 
by M. Markevitch 
\footnote{http://asc.harvard.edu/cal/Hrma/
hrma/optaxis/platescale/geom\_public.html} ).
During the first  observation, the physical distances of the spot centers
from the pulsar were $0.20\pm 0.01\, d_5$~pc and $0.34\pm0.01\, d_5$~pc, where
$d_5 = d/(5\,{\rm kpc})$. 
By the second observation the apparent centers had moved significantly, 
$\sim 3.49^{\prime \prime}\pm0.31^{\prime \prime}$ and 
$\sim 1.90^{\prime \prime}\pm0.44^{\prime \prime}$ 
away from the pulsar, implying apparent transverse velocities
of $\sim 1.4d_5\, c$ and $\sim 0.8d_5\, c$. 



\subsection{Spectral Analysis}

\subsubsection{Methodology}

$Chandra$'s high resolution imaging capabilities allow the measurement
of point source spectra to be made with remarkably low background levels. 
We initially used a 7 pixel radius circular region to extract a pulse height 
spectrum from the pulsar with an annular background region from 7 to 10 
pixels. The spectrum showed an enhancement
at high energies, a clear signature of pileup. From the
observed count rate of $\sim8\times 10^{-2} {\rm cts}\, {\rm s}^{-1}$,
we estimate a pile-up fraction $\sim 5$\%, which is enough to significantly 
affect the measurement of the spectral index. Around half of the
counts are contained within one pixel at the center of the PSF, which
is where most of the pile-up occurs. By excluding the
central pixel, we can minimize the effects of pileup, but must
then worry about the variations in the PSF as a function of energy which
 will affect the spectrum
when a restricted extraction region is used. Since the current version
of CIAO is not able to calculate the effective area when only a fractional
part of the PSF is enclosed, we examined the PSF enclosed energy
curves and carefully chose annular extraction regions for the on source and
background spectra whose difference would provide a relatively
constant enclosed energy fraction over most of the {\it Chandra} waveband.
Choosing an annular source extraction region with inner and outer
radii of $0.3^{\prime \prime}-1.0^{\prime \prime}$ , and a background annular region of
$1.0^{\prime \prime}-1.38^{\prime \prime}$, 
we obtained an enclosed energy fraction of $\sim 0.37$
which varies by only a few percent in the 1-9 keV range.
We fit the resulting spectrum to 
an absorbed power-law model, the results of which are listed in
Table~\ref{specfits}. 

For extended sources, both the cosmic X-ray and particle background
are non-negligible, and vary with position on the detector. 
At the moment, there is no consensus on the best way of approaching
this problem. We have therefore used the following procedure,
based on one developed by \citet{rrk01} for analysis of $ASCA$ GIS data.
We assume that the background consists of three components: a 
particle component
which is not necessarily spatially correlated with the telescope effective area,
an extragalactic component which suffers from absorption by the
Milky Way, and a Galactic component. We further assume that the particle
background is
relatively time independent once the data are screened for background 
flaring events and that the 
two cosmic contaminating components uniformly illuminate the telescope over the S3 chip
field of view. The $Chandra$ science support center has made available
blank-sky observations of high Galactic latitude fields which we assume
accurately represent the particle and unabsorbed extra-Galactic components
of the background.

Using the above assumptions, we account for the different background 
contributions in the following way.
We extract spectra from the region of interest and a 
source-free region of the chip. From each of these, we subtract spectra 
extracted from  blank-sky data sets from the same chip regions in order to
remove the particle background contribution.
Since the
effective exposure varies as a function of chip position, we 
scale the local background spectra by the ratio of average exposures of
the source and background regions using the exposure maps which were
used in producing the exposure corrected images of Paper I. 
Although this corrects for the gross differences in effective area
between the source and background extraction regions, it should
be noted that the variation in effective area with position is energy
dependent, which is not taken into account with the current procedure.
However, this assumption allows us to account for the background
in a model independent way. 
In practice, for G11.2$-$0.3 the spectral results are not strongly
sensitive to the background subtraction method used, as the source
to background ratio is high. 
To create effective area and response matrix files, we make a binned
image of the source region photons from which we subtract an
image created from the blank-sky data set, and use this as the
input weighting map to the CIAO {\tt mkwarf} and {\tt mkrmf} tools.
The resulting source and background spectra and calibration files are
used in XSPEC to fit models to the spectra. 
Note that, given the assumptions described
above, all three background components are accounted for by this
procedure.  

\subsubsection{Results of Model Fitting}

\begin{figure}
\plotone{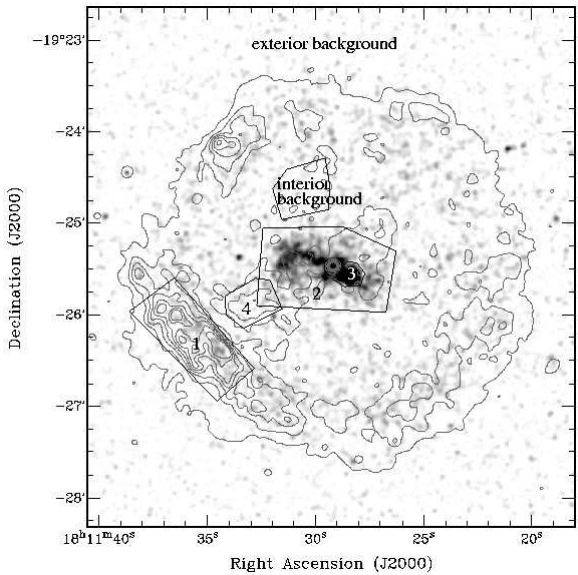}
\figcaption{\label{fig:ximage}
Hard (4--9 keV) X-ray image with soft (0.6--1.65~keV) X-ray
photon contours, showing spectral extraction regions.}

\end{figure}

In Figure~\ref{fig:ximage}, we show the extraction regions used for 
spectral analysis. The greyscale again shows the hard
X-ray emission,  while the
contours show the soft X-ray emission. 
The interior of the shell has enhanced emission
over the general background, so for features within the shell, we choose
a background region interior to the shell.
In the case of the shell itself (region 1), a region exterior to the shell
is chosen as background. 

We analyzed the bright region of the shell (region 1) near that of a portion of the 
interior soft feature (region 4). In order to compare the two regions,
we initially used the constant temperature, plane-parallel shock
plasma model PSHOCK in XSPEC for both. 
However, there is a clear hard excess
in region 1 seen both in the spectrum and the hard X-ray image
(Figure~\ref{fig:ximage}), while there is very little emission above
4 keV in region 4. This hard X-ray emission is expected
from shock acceleration of particles in SNR shells, the presumed
source of Galactic cosmic rays. To account for this, we add 
the synchrotron roll-off SRCUT model \citep{rk99}, which simulates
the roll-off of the synchrotron emission spectrum of an
exponentially cut off power-law distribution of electrons, 
to the region 1 fit, fixing
the radio spectral index $\alpha=0.56$ as determined in  
Paper II and the 1~GHz flux extrapolated from the 1465~MHz image. We then fit 
for the roll-off break frequency which will reproduce 
the observed hard excess.
In both regions, the spectral features seemed to be shifted by $\sim 1.5$
channels. Since we used a 10 eV/channel binning scheme when generating the 
response
matrices, this is similar to the +16 eV zero point gain shift reported on
31 Jan. 2002 to the Chandra users 
committee\footnote{http://asc.harvard.edu/udocs/ucom.jan02.html}. We 
corrected for this by
applying a 1.6 channel bin shift to the response matrix.
A significant improvement to the region 4 fit could be made by adding a weak 
narrow line at 
1.18 keV, and there was weaker evidence for this feature in the other spectra.
The strength of this line  varied with the
size of the extraction region, and not the total flux. We therefore suspect 
that this is some
background component not adequately removed by the procedure described
in the above section.  

For region 4, (and subsequently regions 2 and 3),
we were able to obtain a reasonable fit
using the simple absorption and shock models, but found the much greater
signal to noise region 1 spectrum, with $\sim 25,000$ counts,
showed significant residuals around all
of the spectral line features. Using the VPSHOCK model, which allows
the individual element
abundances to vary, greatly improved the fit, but there still seemed to 
be systematic offsets in the location of the spectral features. Allowing
a redshift improved the fit statistically, but the implied velocities 
were too large to be plausible. We then tried allowing variable
abundances in the absorption using the VPHABS model, 
which  significantly improved the overall fit resulting in a reduced 
$\chi^2\sim1$. However,
there is significant correlation between the individual element 
abundances as well
as between the abundances in the shock model and their corresponding absorption
abundances. It is also true that calibration uncertainties in the response, 
which are known to lead to overestimates of line widths, will have had an
unknown effect on the abundance values.
We therefore do not report the individual abundance values quantitatively
and instead make only general comments. 
It should be noted that the various
model changes only slightly alter the fit temperature and overall 
absorption column of the region 1 spectrum. 
The VPSHOCK and VPHABS also improved the
region 4 spectral fit, but the freedom allowed by poorer statistics (from only
$\sim 4000$ counts in the spectrum)
could not effectively constrain the parameters, and so we only report the
results of the fits without individually varying abundances.

The resulting temperatures of regions 1 and 4 were similar, with the
shell having abundances several times solar.
However, these are all sensitive to the
fit absorption abundances, which tended to be high in Si and S, but
low in Mg. High resolution spectra very sensitive to
low energies would be required to disentangle the
absorption and emission features.
There were several local minima in the
multi-parameter $\chi^{2}$ space near the best fit values,
with moderately different temperatures (ranging from $kT=0.5-0.6$ keV) 
and absorptions. The ionization timescale upper limit $\tau_u$ parameter of
the shock model 
and the SRCUT break frequency
were somewhat more sensitive, but did not vary by more than a factor of 2. 
We emphasize that the 90\% multi-parameter
confidence regions presented in Table~\ref{specfits} are statistical around 
our best fit values and may not
represent the true uncertainty. However, the systematic
uncertainties are not likely to affect the qualitative interpretation
of the results presented in the discussion below. 

We also tried fitting the shell portion using the VNPSHOCK model
which allows for different ion and electron temperatures. There were
no significant differences between the two temperatures, indicating
that the ions and electrons are in rough equilibrium. However,
the dependence of the model on changes in ion temperature is weak \citep{blr01},
and moderate temperature differences could not be ruled out. 

\begin{figure}
\plotone{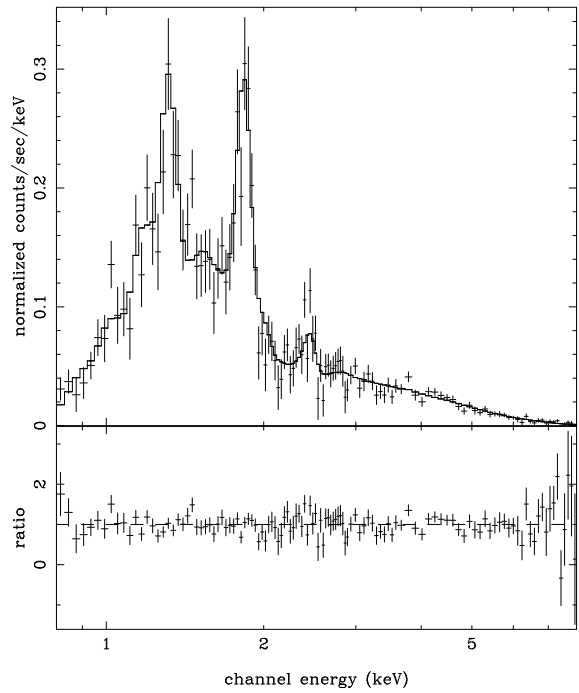}
\figcaption{\label{fig:spec}
ACIS S3 spectrum of PWN (region 2), showing residuals to the 
absorbed PSHOCK + power-law model. The Mg, SI, and S 
line emission from the contaminating thermal emission are
clearly evident, as is the high energy power-law tail from the 
synchrotron nebula. 
  }

\end{figure}

In order to determine the synchrotron emission properties of the
hard PWN (regions 2 and 3), 
the soft thermal emission in the central regions near the pulsar 
needs to be accounted for.
We use our best fit PSHOCK and PHABS values from the interior soft emission
(region 4) spectral fits, and allow only the normalization and total
$n_H$ to vary.  To this we add a simple power-law model to fit the hard PWN 
components. 
The synchrotron X-rays are clearly seen in the data (Figure~\ref{fig:spec}).
The results of all the fits are listed in 
Table~\ref{specfits} with 90\% multi-parameter
confidence regions in parentheses. The observed fluxes with 1$\sigma$ errors
are listed for the energy ranges indicated for the pulsar and regions
1 and 4. For regions 2 and 3, the unabsorbed fluxes of the power-law 
model component only are given. 

\section{Discussion}

Models of PWN evolution in SNRs suggest there are two major epochs
in the life of the system.
In the early evolution, the PWN expands into the
freely expanding ejecta of the SNR. The highly relativistic
pulsar wind is terminated at a shock near the pulsar, which
continuously injects electrons and ions having a power-law distribution of 
energies, as well as magnetic field,
into a surrounding cavity.  The particles then gyrate in the nebular magnetic 
field, causing the synchrotron emission. 
This cavity itself expands at supersonic
speeds relative to the surrounding SNR ejecta. Therefore, there should
be a forward shock at the edge of this cavity, similar to the
SNR forward shock moving into the surrounding ISM. 

As the SNR shell expands, sweeping
up more of the interstellar medium, a reverse shock is launched
which travels towards the SNR center, carrying information about the
external medium to the inner portions of the remnant. When the reverse
shock encounters the forward shock of the PWN, the PWN is compressed,
and reverberates several times \citep{vand01}. 
The compressed PWN finally settles
into a smaller size relative to the SNR shell, and begins to
expand again, this time sub-sonically. 
This corresponds to the SNR entering the Sedov
expansion phase, where the dynamics are dominated by the swept-up
ISM mass. 

Hydrodynamical models of spherically symmetric
PWNe in SNRs suggest the
collision between the reverse shock and the PWN occurs after
$\sim 2000$ yr \citep{vand01,bcf01}. Although the energy input of the 
pulsar in these models was appropriate for the high
energy output of the Crab and not the relatively steady low energy output of
PSR J1811$-$1925, the time of collision is largely determined by the propagation
of the reverse shock which only depends on the
dynamics of the supernova ejecta. We therefore expect the timescale
in the case of G11.2$-$0.3 to be roughly similar.
If we assume the age of the SNR to be 1615 yr, 
corresponding to the 386 A.D. historical association, then we would
expect the time of collision to be near. If the
collision has not yet occurred,  the PWN should be near
its largest extent relative to the SNR shell, and its morphology
should be largely determined by the geometry of the pulsar wind. 
If the collision has occured or is currently occurring, then the 
PWN should be compressed and distorted by the reverse shock. 

The separation of the PWN from the surrounding shell
in both the X-ray and radio images suggests that the reverse shock has not
yet reached the PWN. The hard X-ray emission concentrated at the
inner edge of the shell may result from the recent
passage of the reverse shock, suggesting the latter is still near the
observed shell. This hypothesis is somewhat supported by the 
rather large size of the radio PWN (Paper II).
In this case, one might hope 
to see evidence of the PWN forward shock in the form of thermal X-rays. 
The thermal emission interior to the shell, which we refer to as the soft 
``PWN", may be a result of this shock. The smaller ionization timescale 
of this feature compared to the nearby shell emission may support
this interpretation. The similar temperatures 
suggest the shock speeds are nearly identical, which might be 
unexpected if the radius of the soft ``PWN" is only 2/3 the radius of the 
shell, as it appears in the image. However, the SNR forward shock will
have been significantly slowed by the accumulation of ISM material
over the lifetime of the remnant, and the relative magnitude
of the velocity jumps at the two shock fronts is model dependent
\citep{vand01}. It is therefore difficult to say what relative shock
velocities, and hence relative temperature, should be expected. 

Following the reasoning of \citet{rey94}, the 
Rankine-Hugoniot shock speed can be inferred from the
best fit temperature of $kT\sim 0.58$~keV of the shell region to be 
$v_s=(16 kT/3\mu m_p)^{1/2} \simeq 700 \,{\rm km}\,{\rm s}^{-1}$
assuming electron-ion
equilibrium and that the mean mass per particle $\mu m_p$ is 0.6 times the
proton mass. Since the ions heat the electrons, the observed
electron temperature is a lower bound on the ion temperature
and hence the shock velocity. However, the results of the
VNPSHOCK model suggest that the electrons and ions are not
far from equilibrium. The velocity thus derived
implies a current expansion rate 
of $\dot \theta \la 0.03^{\prime \prime}/d_5 {\rm yr}^{-1}$,
giving
a free expansion age of $\tau_{fe}=R_{snr}/v_s\la 5000d_5$~yr
for the SNR radius $R_{snr}\sim 150^{\prime \prime}$.  
If we adopt the Sedov relation for a blast wave of radius $R_S=2.5v_s \tau$
we obtain a Sedov age estimate of $\tau_S\la 2000d_5$ yr, close to the
historical age $\tau=1615$~yr.
The ionization timescale parameter from the VPSHOCK model 
implies a reasonable electron density of 
$n_{es}\sim 10\, {\rm cm}^{-3}$ for a
1615~yr old shock, while the synchrotron roll-off break
frequency implies a maximum electron energy of $\sim 20$ TeV for a
typical magnetic field strength $B\sim 10\,\mu$G \citep{rk99}. 

The pulsar has a hard spectrum with no sign of a thermal
component, although that is not surprising given the high level of absorption. 
The photon index $\Gamma\sim 1.3$ is similar to what has been
found recently from other Chandra observations of young pulsars 
\citep{l02,p01,gaen02}.
The unabsorbed flux, $F_x\sim 4.2\times 10^{-12} {\rm ergs}\,{\rm cm}^{-2}
\, {\rm s}^{-1}$ in the 1-10 keV energy band, is similar to 
the $\sim 4\times 10^{-12} {\rm ergs}\,{\rm cm}^{-2} 
\, {\rm s}^{-1}$ quoted by \citet{tori97} for the total pulsed flux,
suggesting that the pulsed fraction is nearly 100\%. 

\subsection{Spots as Wisps}

The spatial variability observed in the bright PWN spots 
may be the equivalent of the Crab's wisps, which are seen to form 
at the radius of the inner ring ($r \sim 0.14$~pc) then move outwards
at a speed $v_w\sim 0.5c$ \citep{mori01}. 
In the Gallant and Arons (1994)
model, the wisp spacing roughly corresponds to the post-shock ion Larmor radius.
In such a scenario, the spots should form at the characteristic radii
and then propagate outward and dissipate. The coincident enhanced radio emission
(Figure~\ref{fig:wisps}), 
with its much longer cooling timescale, suggests that these spots are 
persistently found in this region. We also note that there appears to 
be dynamical 
activity within the narrow feature on other side of the pulsar, 
but the variations in the flux levels
are too near the noise level to unambiguously identify individual knots
at both epochs.
The bright features may also represent a collimated, jet-like
outflow. In this case,  we might expect new knots to be ejected near the 
pulsar and move outwards and diffuse. The apparent velocity 
of $c$ or even greater and the brightness asymmetry  around
the pulsar suggest this may be the case.

The asymmetry of the radio emission compared with the X-ray
emission is puzzling. On one side of the pulsar, the emission
is coincident, while on the other, the radio and X-ray emission are
anti-coincident. Such correlations have been observed in other
PWN, most notably around PSR B1509$-$58 \citep{gaen02} whose characteristic
age is similar to the  age of G11.2$-$0.3. 
The central location and corresponding upper
limits on the velocity of the pulsar (see Paper I) make interpretations
of such features as a trail marking the
previous passage of the pulsar through the region untenable, and
argue against a large asymmetry in the birth explosion as
a causal factor. 

\subsection{Magnetic Field}

In this section, we present a very simplistic modification
to standard PWN magnetic field estimates to account for non-sphericity.
We can identify the location of spot 1, $r_{s1}$, as being downstream from
the pulsar wind termination shock $r_{s1}\ga r_t$, and use this to estimate 
the downstream magnetic 
field $B_t$ of the PWN. If we assume a radial relativistic flow and a
postshock velocity of $c/3$, then (eg. Chevalier 2000):

\begin{equation}
B_t = \left( {6\epsilon_t \dot E \over f_t r_t^2 c}\right)^{1/2} \sim \,6\times
10^{-5} \left( {\epsilon_t^{1/2} \over f_t (r_t/r_{s1}) d_5}\right) \, {\rm G},
\end{equation}

\noindent
where $\epsilon_t \la 1$ is the ratio of magnetic energy to total energy
in the postshock flow. To allow for a non-spherical outflow, we include
a beaming fraction at the termination shock $f_t \le 1$.

The current best estimate of the radio energy spectral index is 
$\alpha_r\simeq 0.25$ (defined as $S_{\nu} \propto \nu^{-\alpha_r}$) 
with an acceptable range of $\alpha_r=0.15$-$0.3$. 
The corresponding
range of the X-ray energy index is $\alpha_X=\Gamma - 1=(0.54$-$0.90)$ 
with a best fit
value $\alpha_X=0.73$. The difference between the 
X-ray and radio spectral indices $\delta\alpha_{rX} \sim 0.5$ is 
consistent with what is expected for a single synchrotron cooling 
break between
the radio and X-ray bands, although the uncertainty may be as large as 0.25. 
By assuming $\delta\alpha_{rX}=0.5$, we
can find the break frequency from the measured radio index and
radio and X-ray flux densities:
 
\begin{equation}
\nu_b=\left( {S_X\nu_X^{\alpha_r +0.5} \over S_r\nu_r^{\alpha_r}}\right)^2 .
\end{equation}

\noindent
We estimate the flux density of the radio nebula at 1.5 GHz to be
$S_r(PWN)\sim 0.4$~Jy and of the spot region to be $S_r(spot)\sim 0.04$~Jy,
although these could be off by a factor of two (see Paper II for
a discussion of flux uncertainties).  
For the observed range of values for $\alpha_r$, this results in estimated
upper limits on the break frequencies of $\nu_b(PWN)\la 50$ GHz and $\nu_b(spot)
\la 200$ GHz, with values for the nominal spectral index 
$\alpha_r=0.25$ of $\sim 8$ GHz and $\sim 25$ GHz, respectively.

If we assume constant energy injection of particles over the
lifetime of the PWN, as indicated by the inferred initial spin being nearly
the same as the current spin period (Paper I), we can estimate
the average magnetic field from the apparent break frequency (eg. Chevalier
2000) from:

\begin{equation}
B_N \sim {4\times 10^{-3}  \over 
(\nu_b/10\, {\rm GHz})^{1/3}(t/1614\, {\rm yr})^{2/3}}\, {\rm G} .
\end{equation}

\noindent
This is very large compared to the post-shock estimate above (Eq. 1). 
Even with the most extreme values of
flux and spectral index, because of the weak dependence
on $\nu_b$, the magnetic field in the nebula and in the spot region 
$B_N\ga 10^{-3}$ G. If the magnetic
field is constant throughout the emitting volume of the PWN, then the
maximum magnetic field can be inferred from the energy
injected over the lifetime of the pulsar assuming negligible radiative 
losses (eg. Pacini and Salvati 1973):

\begin{equation}
B_N \le \left( {6 \epsilon_N\dot E t}\over {f_N r_N^3} \right)^{1/2} 
\simeq 2.6 \times 10^{-4} \left({\epsilon_N} \over {f_N d_5^3} \right)^{1/2} 
{\rm G},
\end{equation}

\noindent 
where $\epsilon_N$ is the fraction of the spin-down energy which ultimately 
goes into 
magnetic field (0.5 for equipartition) and $f_N$ is the ratio of the emitting 
volume to a sphere of radius $r_N\sim 40^{\prime \prime}$, the distance 
from the pulsar to the radio arc.  
In order to reconcile Equations 3 and 4, we need a nebular filling fraction
$f_N \sim 0.01$. This is feasible if the emission region is bipolar or toroidal
rather than spherical, and if the region of high magnetic field, where most
of the emission takes place, is 
somehow confined to  the
narrow region of high X-ray emission and the surface of the wind
bubble. 
The observed PWN structure in the 3.5~cm image may support this
picture (see Figure~\ref{fig:wisps}). In general, $f_t \la f_N$,
since we expect that the outflow would not become much further collimated
after the termination shock, but rather  the solid angle
of the flow  would tend to increase with radius. This suggests that the 
pulsar wind is initially
confined to narrow, polar outflows with opening angles $\theta\la 10^{\circ}$
in the case of a bipolar nebula
or a thin, equatorial sheet in the case of a toroidal nebula. 

We introduce the parameters $\beta \equiv (B_N/B_t)^2(\epsilon_t/\epsilon_N)$
and $f\equiv f_t/f_N \la 1$.  It can be shown that the arguments of
\citet{rg74} (see also Kennel \& Coroniti 1984) for the build up of the 
magnetic field to approximate equipartition strength 
may be applied to both a conical and nozzle type flow, in which case we
expect $\beta \sim 1$. 
We can then combine Equations 1 and 4 to derive the following
estimate of the termination shock radius:

\begin{equation}
{r_N\over r_t}  \simeq \left( {fct}\over {\beta r_N}\right)^{1/2} \sim 20,
\end{equation}

\noindent
suggesting $r_t$ should be a few arcseconds from the pulsar.
In that case, monitoring of the emission by {\it Chandra} should resolve the
region of spot formation.

Bandiera, Pacini, and Salvati (1996) noted that the extrapolated {\it ASCA}
PWN spectra combined with previous estimates of the radio PWN flux
led to an anomolously high magnetic field, and suggested a period
of rapid spin-down early in the pulsar's history where the
magnetic field was dumped into the surrounding region, and confined
by the external SNR shell. This was previous to the discovery of
pulsations, and their models greatly overestimate the pulsar spin period.
It is interesting to note, however, that an initial spin period
$P_0 \sim 20$~ms, similar to that of the Crab pulsar's, 
would provide the necessary
energy for the magnetic field. Nevertheless, the arguments made above for
non-sphericity in the nebula obviate the need for an ad-hoc invocation
of an early, short-term, very rapid spin-down era.

Inferring the magnetic field by extrapolating the total PWN radio and X-ray 
spectra assuming a single break may be questionable.
There may be a change in spectral index over the PWN to which we are not
sensitive. There is also a possibility that the 
reverse shock has begun to encounter the PWN, temporarily enhancing the
magnetic field at the edges. 
However, the spot region shows evidence of current particle injection 
and is likely not yet affected by the reverse shock. The
break frequency of this region is $\sim 5$ times higher here than for 
the PWN as a whole, which only lowers the magnetic field estimate
by a factor of 2, assuming this feature has existed for the entire
lifetime of the PWN. 

\section{Conclusions}

We have measured the X-ray spectrum of various components of
the shell and pulsar wind nebula of G11.2$-$0.3 and compared them
with the radio spectrum and morphology. The shell temperature 
and hard X-ray structure are consistent with a $\sim 2000$ yr 
old remnant nearing the onset of the Sedov phase. The pulsar
wind nebula itself shows evidence of relativistic dynamic evolution
of bright X-ray enhancements near the pulsar. We argue that the
reverse shock has not yet reached the expanding pulsar wind bubble, 
and therefore the observed structure is largely determined by the
structure of the wind itself. The morphology and magnetic
field suggested by the low spectral break energy suggests the
outflowing wind is highly non-spherical. The wind
may be largely constrained to a narrow, bipolar outflow, or possibly a
thin, equatorial sheet. Further high-resolution observations at X-ray, radio and
infrared wavelengths could identify
the region of spot formation, help determine the true morphology,
and further constrain the break energy of the spectrum.

\section*{Acknowledgements}

We would like to thank S. Reynolds and  K. Borkowski for useful
discussions on the spectral modelling. 
This work was supported in part by 
$Chandra$ grant GO0-1132X from the Smithsonian Astrophysical Observatory,
NASA LTSA grants NAG5-8063 (VMK) and NAG5-7935 (EVG),
a Quebec Merit Fellowship (MSER), and NSERC research grant RGPIN228738-00 
(VMK). The National Radio Astronomy Observatory is a 
facility of the National Science Foundation operated under cooperative 
agreement by Associated Universities, Inc. 


\begin{thebibliography}{}
\bibitem[Bandiera, Pacini, \& Salvati(1996)]{bps96} Bandiera, R.,
Pacini, F., \& Salvati, M.\ 1996, \apjl, 465, L39
\bibitem[Becker, Markert, \& Donahue(1985)]{bmd85} Becker, 
R.~H., Markert, T., \& Donahue, M.\ 1985, \apj, 296, 461 
\bibitem[Blondin, Chevalier, \& Frierson(2001)]{bcf01} 
Blondin, J.~M., Chevalier, R.~A., \& Frierson, D.~M.\ 2001, \apj, 563, 806 
\bibitem[Borkowski, Lyerly, \& Reynolds(2001)]{blr01} 
Borkowski, K.~J., Lyerly, W.~J., \& Reynolds, S.~P.\ 2001, \apj, 548, 820
\bibitem[Chevalier(2000)]{chev00} Chevalier, R. 2000, \apj, 539, L45
\bibitem[Clark \& Stephenson(1977)]{cs77} Clark, D.~H.~\&
Stephenson, F.~R.\ 1977, Oxford [Eng.] ; New York : Pergamon Press,
1977.~1st ed.
\bibitem[Downes(1984)]{down84} Downes, A.\ 1984, \mnras, 210, 845
\bibitem[Gaensler et al.(2002)]{gaen02} Gaensler, B.~M., 
Arons, J., Kaspi, V.~M., Pivovaroff, M.~J., Kawai, N., \& Tamura, K.\ 2002, 
\apj, 569, 878 
\bibitem[Gallant \& Arons(1994)]{ga94} Gallant, Y. A. \& Arons, J. 1994, \apj, 
435, 230
\bibitem[Gotthelf(2001)]{got01} Gotthelf, E.~V.\ 2001, 20th 
Texas Symposium on relativistic astrophysics, 513
\bibitem[Green et al.(1988)]{gree88} Green, D.~A., Gull, S.~F., Tan,
S.~M., \& Simon, A.~J.~B.\ 1988, \mnras, 231, 735
\bibitem[Helfand, Gotthelf, \& Halpern(2001)]{hgh01} Helfand, D., Gotthelf, 
E., Halpern, J. 2001, ApJ, 556, 380
\bibitem[Kaspi et al.(2001)]{kasp01} Kaspi, V.~M., Roberts, M.~E.,
Vasisht, G., Gotthelf, E.~V., Pivovaroff, M., \& Kawai, N.\ 2001,
\apj, 560, 371 (Paper I)
\bibitem[Kaspi \& Helfand(2002)]{kh02} Kaspi, V.~M.~\& 
Helfand, D.~J.\ 2002, To appear in ``Neutron Stars in Supernova 
Remnants" (ASP Conference Proceedings), eds P.~O.~Slane and B.~M.~Gaensler 
\bibitem[Kennel \& Coroniti(1984)]{kc84} Kennel, C. F. \& Coroniti, F. V. 
1984, \apj, 283, 694
\bibitem[Kothes \& Reich(2001)]{kr01} Kothes, R.~\& Reich, W.\ 2001,
\aap, 372, 627
\bibitem[Lu et al.(2002)]{l02} Lu, F.~J., Wang, Q.~D., 
Aschenbach, B., Durouchoux, P., \& Song, L.~M.\ 2002, \apjl, 568, L49 
\bibitem[Mori et al. (2001)]{mori01} Mori, K., Hester, J.~J.,
Burrows, D.~N., Pavlov, G.~G., \& Tsunemi, H.\ 2001,To
appear in ``Neutron Stars in Supernova Remnants" (ASP Conference
Proceedings), eds P.~O.~Slane and B.~M.~Gaensler.
\bibitem[Morsi \& Reich(1987)]{mr87} Morsi, H.~W.~\& Reich, W.\
1987, \aaps, 71, 189
\bibitem[Pacini \& Salvati (1973)]{ps73} Pacini, F. \& Salvati, M.\ 1973,
\apj, 186, 249
\bibitem[Pavlov et al.(2001)]{p01} Pavlov, G.~G., Zavlin, 
V.~E., Sanwal, D., Burwitz, V., \& Garmire, G.~P.\ 2001, \apjl, 552, L129
\bibitem[Pavlov et al.(2001)]{pav01} Pavlov, G.~G., Kargaltsev, O.~Y., Sanwal, 
D., \& Garmire, G.~P.\ 2001, ApJL, 554, L189
\bibitem[Rees \& Gunn(1974)]{rg74} Rees, M.~J.~\& Gunn, 
J.~E.\ 1974, \mnras, 167, 1
\bibitem[Reynolds \& Chevalier(1984)]{rc84} Reynolds, S.~P.~\&
Chevalier, R.~A.\ 1984, \apj, 278, 630
\bibitem[Reynolds \& Keohane(1999)]{rk99} Reynolds, S.~P.~\& 
Keohane, J.~W.\ 1999, \apj, 525, 368
\bibitem[Reynolds et al.(1994)]{rey94} Reynolds, S.~P., Lyutikov, M.,
Blandford, R.~D., \& Seward, F.~D. 1994, \mnras, 271, L1
\bibitem[Roberts, Romani \& Kawai(2001)]{rrk01}Roberts, M., Romani, R. \& 
Kawai, N.\ 2001, \apjs, 133, 451
\bibitem[Scargle(1969)]{scar69} Scargle, J. D. 1969, \apj, 156, 401
\bibitem[Tam, Roberts, \& Kaspi(2002)]{trk02} Tam, C., 
Roberts, M.~S.~E., \& Kaspi, V.~M.\ 2002, \apj, 572, 202 (Paper II)
\bibitem[Torii et al.(1997)]{tori97} Torii, K., Tsunemi, H., Dotani,
T.,~\& Mitsuda, K.\ 1997, \apjl, 489, L145
\bibitem[Torii et al.(1999)]{tori99} Torii, K., Tsunemi, H., Dotani,
T., Mitsuda, K., Kawai, N., Kinugasa, K., Saito, Y.,~\& Shibata, S.\
1999, \apjl, 523, L69
\bibitem[van der Swaluw et al.(2001)]{vand01} van der Swaluw, E.,
Achterberg, A., Gallant, Y.~A.~\& T\'{o}th, G.\ 2001, \aap, 380, 309
\bibitem[Vasisht et al.(1996)]{vasi96} Vasisht, G., Aoki, T., Dotani,
T., Kulkarni, S.~R., \& Nagase, F.\ 1996, \apjl, 456, L59
\bibitem[Weisskopf et al.(2000)]{wei00} Weisskopf, M. et al. 2000, ApJ, 536, L81

\end{thebibliography}

\clearpage

\clearpage

\begin{deluxetable}{lccccccccc}
\tabletypesize{\scriptsize}
\tablecaption{Spectral fits \label{specfits}}
\tablewidth{0pt}
\rotate
\tablehead{
\colhead{Model} & \colhead{Parameter} & \colhead{Pulsar} & \colhead{Shell (1)} 
& \colhead{PWN (2)} & \colhead{Spots (3)} & \colhead{Soft ``PWN" (4)} 
}
\startdata
(V)PHABS\tablenotemark{a} & $n_H\, (10^{22}\,{\rm cm}^{-2})$& 
2.36(1.75-3.19)\tablenotemark{b}
& 1.71(1.70-2.36) & 2.14(2.04-2.25) & 2.28(1.86-2.77) & 2.15(2.03-2.27) \\
POWER\tablenotemark{c} & $\Gamma$ & 1.11(1.00-1.48)& -- & 1.73(1.54-1.90)&1.73(1.55-1.95) &-- \\
(V)PSHOCK\tablenotemark{d} & $kT\,$(keV) & --&0.581(0.553-0.595)& 0.583(frozen)& 0.583(frozen) 
& 0.583(0.517-0.691) \\
& $\tau_u\, (10^{11}\,{\rm s}\,{\rm cm}^{-3})$& --&6.7(5.7-8.1)& 
4.2(frozen)& 4.2(frozen) & 4.2(2.3-14.9)\\
SRCUT\tablenotemark{e} & $\alpha$& --& 0.56& --& --&--\\
& $\nu_b\,(10^7\,$~GHz) & --& 1.80(1.68-1.90)& --& --& --\\
& $S_{1GHz}\,$(Jy) & --& 2.0& --& --&--\\
\multicolumn{2}{c}{$F_X\,(10^{-12}\,{\rm ergs}\,{\rm cm}^{-2}\,{\rm s}^{-1})$} 
& $3.29\pm0.14$ (1-10 keV)& $3.70\pm 0.03$ (0.7-8 keV) & $4.44\pm0.12$\tablenotemark{f} (1-10 keV) & $0.87\pm0.04
$\tablenotemark{f} (1-10 keV)  & $0.37\pm0.01$ (0.7-8 keV)\\
 \enddata
\tablenotetext{a}{Photoelectric absorption model, 
allowing the elemental abundances to vary for the shell (1) region only.}
\tablenotetext{b}{Values in parentheses are 90\% multi-parameter confidence regions.} 
\tablenotetext{c}{simple photon power law}
\tablenotetext{d}{Plane Parallel shock model of \citet{blr01}, 
allowing the elemental abundances to vary for the shell (1) region only.}
\tablenotetext{e}{Synchrotron roll-off model of \citet{rk99}.}
\tablenotetext{f}{Unabsorbed flux of power law component only.}

\end{deluxetable}

\end{document}